\documentclass[%
superscriptaddress,
twocolumn,
10pt,
amsmath,amssymb,
aps,
prb,
floatfix,
]{revtex4-1}

\usepackage{graphicx}
\usepackage{dcolumn}
\usepackage{bm}
\usepackage{gensymb}
\usepackage{blindtext}

\usepackage{changes}
\usepackage{lipsum}
\usepackage{caption}
\usepackage{subcaption}

\usepackage{bm}
\usepackage{amsmath}
\usepackage{hyperref} \hypersetup{colorlinks=true,citecolor=blue,linkcolor=blue,urlcolor=blue}
\usepackage{amsmath}
\usepackage{multirow}
\usepackage{booktabs}
\usepackage{epsfig}
\usepackage{xcolor,soul}
\usepackage{epstopdf}
\usepackage{lipsum,adjustbox}
\usepackage{breakcites}

\begin{document}

\preprint{APS/123-QED}

\title{A Lorenz gauge formulation for TDDFT}

\author{Dor Gabay}
\affiliation{Department of Physical Electronics, Tel Aviv University, Tel Aviv 69978, Israel.}
\email{dorgabay@mail.tau.ac.il}
\author{Ali Yilmaz}
\affiliation{Institute for Computational Engineering and Sciences, The University of Texas at Austin, TX, USA.}
\author{Vitaliy Lomakin}
\affiliation{Department of Electrical and Computer Engineering, University of California San Diego, CA, USA.}
\author{Amir Boag}
\affiliation{Department of Physical Electronics, Tel Aviv University, Tel Aviv 69978, Israel.}
\author{Amir Natan}
\affiliation{Department of Physical Electronics, Tel Aviv University, Tel Aviv 69978, Israel.}
\affiliation{The Sackler Center for Computational Molecular and Materials Science, Tel-Aviv University, Tel-Aviv 69978, Israel.}
\email{Corresponding author: amirnatan@tau.ac.il}

\begin{abstract}
We describe the inclusion of electrodynamic fields in Time-Dependent Density Functional Theory (TDDFT) by incorporating both the induced scalar and vector potentials within the time-dependent Kohn-Sham equation. The Hamiltonian is described in both the Coulomb and Lorenz gauges, and the advantages of the latter are outlined. Integral expressions are defined for the retarded potentials of each gauge and a methodological approach to evaluate these nontrivial expressions with low computational cost is adopted. Various molecular structures of relatively small sizes are studied, including water, benzene, and conductive carbon chains. Absorption cross sections resulting from both pulse and boost excitations suggest a preserved gauge-invariance of the proposed formal approach to TDDFT in the weak magnetic field limit. 
\end{abstract}

\pacs{Valid PACS appear here}

\maketitle

\section{\label{sec:level1}Introduction}

The progress of nanoscale devices leads to a growing need for the integration of electromagnetic models with atomistic quantum simulations. Integrating the electromagnetic and quantum realms is vital for the proper characterization of electromagnetic phenomena at the nanoscale. 

Time-dependent Density Functional Theory (TDDFT) reliably predicts the optical and electronic properties of molecules in the presence of external electric fields~\cite{runge1984density,marques2012fundamentals,ullrich2012time}. The unique mapping between the scalar potential and particle density, given an accurately defined exchange-correlation functional, provides the means necessary to bypass the complicated many-body wavefunction of Schr\"odinger's equation. One approach to solve the time-dependent Kohn-Sham equations is the linear response method\cite{ullrich2012time,marques2012fundamentals,casida1995time}. Another approach, is that of direct time propagation, in which the single-electron orbitals are marched forward in time using various propagation schemes~\cite{castro2004excited,castro2018propagators,mundt2009real,marques2012fundamentals,ullrich2012time}. The advantage of linear response methods is that they can get the entire optical response spectrum with a reasonable computational effort. The advantage of real-time propagation methods is that they allow the description of non-linear effects such as high-harmonics and ionization by strong lasers\cite{mundt2007photoelectron,son2009multielectron,telnov2009effects,yahel2018effect,floss2018ab}. In addition, the expression for the exchange correlation potential in the adiabatic approximation can be used directly without the need to develop the exchange correlation kernel.

The traditional assumption of null magnetic fields underlying the proof of the Runge-Gross theorem, which gives the formal basis of TDDFT, is that which dwells within an electrostatic regime. In such a frame of reference, only the scalar portion of the Coulomb gauge is preserved, suppressing the induced magnetic fields and the retardation effects which follow. In accounting for the presence of magnetic fields, time-dependent Current Density Functional Theory (TDCDFT)~\cite{vignale2004mapping} has been introduced as a means of extending the one-to-one mapping of TDDFT to the 4-vector potential and current density. While it is possible to add magnetic field contributions into the TDDFT Hamiltonian, it is impossible to formulate a gauge invariant TDDFT Hamiltonian that can account for strong magnetic fields~\cite{vignale1987density}. 

A successful approach~\cite{yabana2012time} to include retardation effects is that of multi-scale simulations, whereby TDDFT and Maxwell's equations are time propagated simultaneously at small and large length scales, respectively. In a proceeding work, Yamada et al. \textit{et al}.~\cite{yamada2018time} solved Maxwell's differential equations for the vector potential to exemplify the affect of ultrashort pulses on thin films solely in the presence of normally incident waves.

In this work, we study, as an approximation, the inclusion of induced magnetic fields into the time-dependent Kohn-Sham equation (TDKS) using both the Coulomb and Lorenz gauge-fixing conditions. In the limit of weak magnetic fields, such an approximation can be useful to observe classical electromagnetic retardation effects within quantum systems. While the formal gauge invariance is not assured, we surmise that, in the presence of relatively weak magnetic fields, gauge invariance can be attained in practice. 

We therefore foresee three possible regimes of interest - (i) in the limit of strong magnetic fields, one must resort to TDCDFT; (ii) in the presence of weaker magnetic fields, one can attain gauge invariance in practice and, as a result, observe physically meaningful classical electromagnetic retardation effects; (iii) in the limit of very small system sizes, one can expect that retardation effects are negligible and, as a result, not only that gauge invariance exists in practice, but that the traditional TDDFT assumption (of null magnetic field) also coheres to the various gauges. 

In this paper, we demonstrate the realization of regime (iii). Our long term prospect is that the formal approach to be presented suffices also for regime (ii). One practical realization would be to study classical retardation effects associated with the electromagnetic interaction of multiple electronic structures partitioned into domains separated by a fraction of the external electromagnetic field's wavelength. This could result in fascinating structures which could be practically realized for various applications. 

Although the Coulomb gauge is the commonly adopted gauge-fixing condition in TDDFT, the Lorenz gauge is shown to be just as effective in characterizing the response of electronic structures, bypassing the need for a projection scheme of the current density, via Helmholtz decomposition. In a large amount of previous works, retardation effects in the induced potential are seldom considered. One possible reason is that calculating retarded potentials can be computationally costly. One alternative route to solving Maxwell's differential equations is in the use of the equivalent integral expressions. In fully incorporating the scalar and vector potentials, a computational bottleneck exists in evaluating retarded potentials due to their intrinsic dependence on past densities. To overcome this, highly efficient FFT-based integral methods are employed. 

In what follows, we start with a formal introduction to the proposed theoretical framework for the incorporation of retarded potentials for different gauge-fixing conditions. Thereafter, a methodological approach to efficiently computing the 4-vector potential is presented. We then exemplify the formal approach on small molecules such as water, benzene, and cumulene. Finally a discussion is held to interpret the results and explore future prospects for the use of alternative gauges within TDDFT.

\section{Theory}
Runge and Gross~\cite{runge1984density} showed that a unique v-representable particle density $n(\textbf{r},t)$ exists for every pair of a time-dependent external potential $v_{ext}(\textbf{r},t)$ and an initial many-body wavefunction $\Psi_{0}$. The time-dependent Kohn-Sham orbitals $\psi_{i,\sigma}$, characterized by a collinear spin-index $\sigma$, must then obey the single-particle time-dependent Kohn-Sham (KS) equation: 
\begin{eqnarray}
i\frac{\partial}{\partial t}\psi_{i,\sigma}(\textbf{r},t)=\Bigl(-\frac{1}{2}\nabla^2+v_{s,\sigma}[n_{\sigma}]\Bigr)\psi_{i,\sigma}(\textbf{r},t) \label{Hstat}
\end{eqnarray}
\begin{eqnarray}
v_{s,\sigma}[n_{\sigma}]=v_{ext}+v_{ion}+v_{ind}[n]+v_{xc,\sigma}[n_{\sigma}] \label{effec_pot}
\end{eqnarray}
In determining the $i=1,...,N$ single-electron orbitals $\psi_{i,\sigma}(\textbf{r},t)$, Eq.~\ref{Hstat} can either be solved using linear response theories or by propagating the equation of motion forward in time, of which we adopt the latter. $v_{s,\sigma}$ represents the sum of the external $v_{ext}$, ionic $v_{ion}$, exchange-correlation $v_{xc,\sigma}$, and induced (Hartree) $v_{ind}$ scalar potentials. The density-dependent induced scalar potential $v_{ind}[n]$, is defined in an electrostatic manner by solving the static Poisson equation or the equivalent integral:
\begin{align}
&\ \mathbf{\nabla}^2 v_{ind}(\mathbf{r},t)=-4\pi n(\mathbf{r},t) \notag \\
&\Leftrightarrow  v_{ind}(\mathbf{r},t)= \int \frac{n(\mathbf{r^\prime},t)d\mathbf{r^\prime}}{|\mathbf{r}-\mathbf{r^\prime}|}
\end{align}

The inclusion of only scalar potentials within the Kohn-Sham Hamiltonian is known as the length gauge. For small systems, the external potential, $v_{ext}$, is typically defined using the dipole approximation (DA), $v_{ext}=\textbf{r}\cdot \textbf{E}_{ext}$, for a spatially uniform, time-dependent external electric field $\textbf{E}_{ext}$. The DA, adopted in such a manner, holds only to first-order in the multipole expansion and therefore, by virtue of the resulting uniform electric field, disregards any contributions brought about by magnetic fields $\textbf{B}$. Such an approximation suffices for systems perturbed at relatively low frequencies and system sizes (small size-to-wavelength ratios), but fails in the presence of spatially varying electric fields. 
Beyond the DA regime, retardation begins to play a dominant role and the full 4-vector potential must be incorporated into the system Hamiltonian. The inclusion of the vector potential, along with the scalar potential, is better known as the velocity gauge. Irrespective of TDDFT, it has been shown that the velocity and length gauges differ substantially for strong-field photoionization processes~\cite{bauer2008comparison} and for the photodetachment of negative Fluorine ions~\cite{reiss2007velocity} in small molecules. It is therefore important to identify the limits of the length gauge and the formal structure of the velocity gauge. Within the Kohn-Sham approach, the dynamical Hamiltonian $\hat{\textup{H}}_{dyn}$ can be written in the velocity gauge by more closely following the formalism of TDCDFT~\cite{vignale2004mapping}: 
\begin{widetext}
\begin{eqnarray}
\hat{\textup{H}}_{dyn}=\frac{1}{2}\Bigl(i\nabla+\frac{1}{c}(\textbf{A}_{ext}+\textbf{A}_{ind}[\textbf{j}_{p}]+\textbf{A}_{xc,\sigma}[\textbf{j}_{p,\sigma}])\Bigr)^2+v_{ext}+v_{ion}+v_{ind}[n]+v_{xc,\sigma}[n_{\sigma}] \label{Hdyn}
\end{eqnarray}
\end{widetext}
Here, a dynamical contribution has been added to the Hamiltonian of Eq.~\ref{Hstat}. The vector potential includes three terms: the external potential $\textbf{A}_{ext}$, spin-dependent exchange-correlation potential $\textbf{A}_{xc,\sigma}$, and the newly introduced induced vector potential $\textbf{A}_{ind}$. The induced and exchange-correlation vector potentials are both functionals of the paramagnetic current density $\textbf{j}_{p,\sigma}$. The total spin-dependent current density $\textbf{j}_{\sigma}$ contains an additional spin-dependent diamagnetic $\textbf{j}_{d,\sigma}$ contribution to the already present paramagnetic $\textbf{j}_{p,\sigma}$ current density
\begin{eqnarray}
\textbf{j}_{\sigma}&&=\textbf{j}_{p,\sigma}+\textbf{j}_{d,\sigma} \nonumber \\
&&=\sum_{i}\frac{\textit{i}}{2}[\psi_{i,\sigma}^{*}\nabla\psi_{i,\sigma}-(\nabla\psi_{i,\sigma}^{*})\psi_{i,\sigma}]+\frac{1}{c}n_{\sigma}\textbf{A}_{s,\sigma} \label{crnt_dnsty}
\end{eqnarray}
Here, $\textbf{A}_{s,\sigma}$ is the sum of all vector potential contributions. Both paramagnetic $\textbf{j}_{p,\sigma}$ and diamagnetic $\textbf{j}_{d,\sigma}$ current densities are required to properly satisfy the continuity equation $\nabla\cdot \textbf{j}=-\partial{n}/\partial{t}$ (where $\textbf{j}=\sum_{\sigma}\textbf{j}_{\sigma}$). The functional dependence of $\textbf{A}_{s,\sigma}$ solely on $\textbf{j}_{p,\sigma}$, and not $\textbf{j}_{d,\sigma}$, is of fundamental importance~\cite{vignale2004mapping}; and the total contribution should only be considered when the full dynamics of the current density are required. 

Within the velocity gauge, the preferred electromagnetic gauge-fixing condition is a matter of choice. In defining the gauge-fixing condition, two particular options of interest are the Coulomb ($\nabla\cdot \textbf{A}_{s}=0$) and Lorenz ($\partial_{t}v_{s}+\frac{1}{c}\nabla\cdot\textbf{A}_{s}=0$) gauges. Brill \textit{et al}.~\cite{brill1967causality} analyzed the equivalence of the two gauges and exemplified their differences in emphasizing the additional projection operation $\textbf{j}\to\mathcal{P}(\textbf{j})=\textbf{j}_T$ which is required in dealing with the Coulomb gauge. The Coulomb gauge is better suited in considering nonrelativistic canonical representations of quantum mechanics. Use of the Lorenz gauge within canonical representations can prove difficult due to either the need for: (1) an additional divergence operation of the vector potential $\textbf{A}_{s}$; (2) an additional time-derivative of the scalar potential $v_{s}$, retrieved via the Lorenz gauge-fixing condition. Although the Coulomb gauge is typically used for TDDFT, we argue that the Lorenz gauge has some major advantages for the following reasons: (a) it does not require a projection scheme to map the 3-vector (i.e. current density) to its transverse field; (b) the Coulomb gauge typically leads to the need for surface integral expressions, via the Helmholtz decomposition~\cite{jackson2001gauge,jackson2002gauge}, to satisfy boundary conditions. Such surface integral expression can be ignored if the current density is small enough at the boundary; (c) The Lorenz gauge naturally articulates the \textit{physical} dynamic behavior of electromagnetic fields. `\textit{Physical}' implies that it is not conceptually luring to characterize time-dependent behavior of electromagnetic fields in an electrostatic manner (without retardation), as is typically done in the time-independent case. 


It is important to emphasize that the proposed electromagnetic excitation is purely classical and within the adiabatic framework; there is no spontaneous emission of any kind. A fully consistent treatment of the exchange-correlation vector potential $\textbf{A}_{xc,\sigma}$ is typically performed within TDCDFT~\cite{vignale1996current}, which we choose to temporarily ignore under the assumption that these effects are negligible in the limit of weak magnetic fields. Whether this assumption is justified becomes apparent if gauge-invariance is still conserved in such a limit. The resulting equation of motion for the single-particle orbitals $\psi_{i,\sigma}$ is:
\begin{equation}
\textit{i}\frac{\partial}{\partial{t}}\psi_{i,\sigma}=\Bigl[\frac{1}{2} \bigl(\textit{i}\nabla + \frac{1}{c}(\textbf{A}_{ext}+\textbf{A}_{ind}[\textbf{j}_{p}])\bigr)^{2} +v_{s,\sigma}[n_{\sigma}]\Bigr] \psi_{i,\sigma} \label{eom_dyn} 
\end{equation}
In applying the Coulomb gauge-fixing condition, the induced scalar $v_{ind}$ and vector $\textbf{A}_{ind}$ potentials take the form 
\begin{eqnarray}
v_{ind}(\textbf{r},t)&&=\int{\frac{n(\textbf{r}',t)}{|\textbf{r}-\textbf{r}'|}d\textbf{r}'} \label{coulomb_gaugeS} \\
\textbf{A}_{ind}(\textbf{r},t)=&&\int{\frac{\textbf{j}_{p,T}(\textbf{r}',t-\frac{|\textbf{r}-\textbf{r}'|}{c})}{|\textbf{r}-\textbf{r}'|}d\textbf{r}'}  \label{coulomb_gauge}
\end{eqnarray}
These expressions can easily be derived using the Coulomb gauge-fixing condition ($\nabla\cdot \textbf{A}_{s}=0$), gauge free relations, and Maxwell's equations (Appendix A). A projection scheme $\textbf{j}_{p}\to \textbf{j}_{p,T}$ is necessary to attain the transverse part of $\textbf{j}_{p}$. 
The projection scheme used to attain the transverse component of the current density $\textbf{j}_{p}$ can more concretely be defined by the Helmholtz decomposition
\begin{eqnarray}
&&\textbf{j}_{p}(\textbf{r})=\frac{1}{4\pi}\nabla^{2}\int{\frac{\textbf{j}_{p}(\textbf{r}')}{|\textbf{r}-\textbf{r}'|}d\textbf{r}'} \nonumber \\
&&=-\frac{1}{4\pi}\Biggl[\nabla\Bigl[\nabla\cdot\int{\frac{\textbf{j}_{p}(\textbf{r}')}{|\textbf{r}-\textbf{r}'|}d\textbf{r}'}\Bigr] + \nabla\times\Bigl[\nabla\times\int{\frac{\textbf{j}_{p}(\textbf{r}')}{|\textbf{r}-\textbf{r}'|}d\textbf{r}'}\Bigr]\Biggr] \nonumber \\
&&=\textbf{j}_{p,L}(\textbf{r})+\textbf{j}_{p,T}(\textbf{r}) \label{helmholtz_v0}
\end{eqnarray}
Here, $\textbf{j}_{p,L}$ is the longitudinal current density and can be shown to have a one-to-one correspondence to the electrostatic potential $v_{ind}$, via the continuity equation
\begin{eqnarray}
\nabla\cdot\textbf{j}&&=-\frac{\partial{n}}{\partial{t}} \nonumber \\
\nabla\Bigl[\int{\frac{\nabla\cdot\textbf{j}(\textbf{r}')}{|\textbf{r}-\textbf{r}'|}d\textbf{r}'}\Bigr]&&=-\frac{\partial}{\partial{t}}\nabla\Bigl[\int{\frac{n(\textbf{r}')}{|\textbf{r}-\textbf{r}'|}d\textbf{r}'}\Bigr] \nonumber \\ 
\textbf{j}_{L}&&=\textbf{j}_{p,L}=-\frac{\partial}{\partial{t}}(\nabla{v_{ind}}) \label{cnty_eqn}
\end{eqnarray}
Given that the vector potential is a purely transverse entity, $\textbf{j}_{d,L}=0$, justifying the presumed equality $\textbf{j}_{L}=\textbf{j}_{p,L}$. Unlike the longitudinal current density, the transverse component $\textbf{j}_{p,T}$ is entirely independent of the density and can more easily be evaluated in its spectral representation
\begin{eqnarray}
\nabla\times\Bigl[\nabla\times\int{\frac{\textbf{j}_{p}(\textbf{r}',t)}{|\textbf{r}-\textbf{r}'|}d\textbf{r}'}\Bigr] \to -\hat{\textbf{k}}\times\hat{\textbf{k}}\times\tilde{\textbf{j}}_{p} \label{jTspectral}
\end{eqnarray}
Here, $\tilde{\textbf{j}}_{p}$ is the Fourier transformed paramagnetic current density and $\hat{\textbf{k}}=\textbf{k}/||\textbf{k}||$ is the normalized coordinate vector in the spectral domain. Applying this projection scheme at every iteration, in propagating the single-particle orbitals in time, is a serious complication which can be overcome by simply adopting a different gauge. In particular, within the Lorenz gauge, retardation is considered in both the scalar and vector potentials, with no need for the aforementioned projection scheme
\begin{eqnarray}
v_{ind}(\textbf{r},t)=\int{\frac{n(\textbf{r}',t-\frac{|\textbf{r}-\textbf{r}'|}{c})}{|\textbf{r}-\textbf{r}'|}d\textbf{r}'} \label{lorentz_gaugeS} \\
\textbf{A}_{ind}(\textbf{r},t)=\int{\frac{\textbf{j}_{p}(\textbf{r}',t-\frac{|\textbf{r}-\textbf{r}'|}{c})}{|\textbf{r}-\textbf{r}'|}d\textbf{r}'}  \label{lorentz_gauge}
\end{eqnarray}
In assuming the Lorenz gauge-fixing condition, the divergence of the vector potential $\nabla\cdot \textbf{A}_{s}$ can no longer be eliminated. A thorough derivation of these integral expressions is partially provided in Appendix B for completeness. Given the potentials intrinsically determine the densities, via the single-particle orbitals, the retarded potentials should naturally subsume the retardation effects within the densities. An interesting prospect then comes forth: in defining the density-dependent potentials, such as that of the exchange-correlation, the history of densities may more easily be incorporated using the proposed formal approach. 
Stepping away from the static characterization of potentials may be necessary to venture beyond the all-too familiar adiabatic regime. However luring, accounting for such a history of densities can be computationally tedious. For this reason, a methodology is proposed in the following section to relieve the complexity of such integral expressions.

\section{Methodology}

Calculating retarded potentials, via Eq.~\ref{coulomb_gauge},\ref{lorentz_gaugeS}-\ref{lorentz_gauge}, is a cumbersome task and fast computational techniques are required to make their evaluation tractable. Their direct evaluation is too computationally prohibitive, particularly because they must be re-evaluated at every time-step in propagating the Kohn-Sham orbitals. For some component $\rho$ of the 4-vector current density, the corresponding induced potential $\phi$ can be efficiently computed by more concretely defining the retarded Green's function within the integral expression
\begin{equation}
\phi(\mathbf{r},t) = \int{g(R,t)*\rho\left(\textbf{r}',t\right) d\mathbf{r^\prime}}
\label{convolution}
\end{equation}
Here, $R=|\textbf{r}-\textbf{r}'|$ is simply the radial separation and will be adopted in the remainder of this section for convenience. The smoothed Green's function $g(R,t)$ takes the form
\begin{equation}
g(R,t)=\left\{
\begin{array}{c l}	
\frac{1}{R}\delta(t-\frac{R}{c}) & R>d\\
& \\
g_{opt}(R)\delta(t-\frac{R}{c}) & R \le d
\end{array}\right.
\end{equation} 
where $d$ is typically a distance of few grid separations, the asterisk denotes temporal convolution, and $g_{opt}$ is the numerically optimized (NOPT) discrete Green's function kernel~\cite{gabay2017optimizing}. The NOPT kernel is simply an analytically approximated Green's function which was numerically optimized at the region of the Coulomb singularity to best approximate the Poisson integral on a grid~\cite{gabay2017optimizing}. The same kernel has been used for the fast evaluation of Poisson integrals and screened Poisson integrals within hybrids and screend hybrids functionals, respectively~\cite{boffi2016efficient,gabay2017size}. 

To evaluate the time dependent potentials, the charge and current densities are sampled uniformly in space using $N_v=N_x\times N_y\times N_z$ grid points separated by $\Delta_x=\Delta_y=\Delta_z=h$ and uniformly in time using $N_t$ time samples with a time-step size of $\Delta t$. The samples are interpolated in time at each grid point $n^\prime=1,\dots,N_v$ using piecewise linear temporal basis functions $T$~\cite{kaur2011accuracy}
\begin{equation}
\rho(\mathbf{r}_{n^\prime},t) \simeq \mathbf{\rho}_0\left[n^\prime\right]+\sum_{l^\prime=1}^{N_t} \rho_{l^\prime}\left[n^\prime\right]T(t-l^\prime \Delta t),
\label{basis_1}
\end{equation}
where $\rho_{l^\prime}$  is a size $N_v$ vector that stores the charge and current density samples at time $l^\prime \Delta t$. The initial density $\rho_0$ simply corresponds to the ground state solution of the electronic structure. 

Eq.~\ref{convolution} is a convolution procedure which respects the causal structure of space-time. Its direct evaluation is expensive, reaching order of $\mathcal{O}(N_{t} N_{v}^{2})$, however, as in the static scenario, it is possible to utilize FFT techniques to reduce the computational complexity. 
Using a simple 3D FFT scheme along each slice of time reduces the order of operations to order $\mathcal{O}(N_{t} N_{g}N_{v}\log{N_{v}})$, where $N_{g}$ is the number of saved past densities. Such an approach typically suffices in the study of small molecules and molecular chains with lengths reaching up to a few nanometers. Any larger electronic structures require more advanced methodologies, such as those taking advantage of the Green's function three-level block-Toeplitz structure~\cite{yilmaz2002fast,hairer1985fast,yilmaz2004time} to apply FFTs along the temporal direction, further reducing the computational complexity to $\mathcal{O}\bigl(N_{t} N_{v}(\log{N_{v}}+\log^{2}N_{g})\bigr)$ (`4D FFT' algorithm).

\section{Results}

We use the Bayreuth version~\cite{mundt2007photoelectron,mundt2009real} real-time propagation TDDFT code as part of the PARSEC real-space package~\cite{chelikowsky1994higher,kronik2006parsec} as a starting point for our code implementation. 

In the examples which follow, we checked our implementation for small molecules such as water, benzene, and an elongated carbyne molecule, C$_{12}$H$_{4}$. We use norm conserving pseudopotentials~\cite{Troullier1991} for carbon, hydrogen, and oxygen with cutoff radii (a.u.) of 1.6/1.6, 1.39, 1.3/1.3 respectively for s/p orbitals throughout. The grid spacing used for the electronic structures to follow is 0.4 a.u. . The single-particle orbitals $\psi_{i,\sigma}$ are propagated using a 4th-order Taylor expansion for the time-evolution operator, $e^{-iH\Delta t}$, with a predictor-corrector scheme to assure proper convergence~\cite{mundt2007photoelectron,mundt2009real}. The adiabatic local density approximation (ALDA) is hereby adopted for all of the exchange-correlation functionals.

Two measures are used in comparing the dynamical behavior of the individual gauges: (1) the dipole moment resulting from sine squared pulses; (2) the dipole moment resulting from a boost excitation in attaining the response of the system. 

In the former (1), laser wavelengths ranging from 800-1600nm are adopted to assure the weak magnetic field limit is satisfied. This guarantees that the external electric field does not dramatically vary along the molecular structure of interest, resulting in a near-null magnetic field. 

In the latter (2), the response is attained using an approach similar to that suggested by Yabana \textit{et al}.~\cite{yabana2006real}: a boost excitation is applied, taking the form of a momentary impulse with a specified amplitude $I$. In both methods, the resulting scalar valued absorption cross section (ACS) $\sigma(\omega)$ is defined by:
\begin{eqnarray}
\sigma(\omega)&=\Biggl|\frac{\pi^{2}\omega}{mcI}\int_{0}^{\infty}dt e^{i\omega t}\overrightarrow{\mathcal{D}}(t)\Biggr|^{2} \label{cross_section} \\
\overrightarrow{\mathcal{D}}(t)&=\int \textbf{r}\sum_{i}|\psi_{i}(\textbf{r},t)|^{2}d\textbf{r} \nonumber
\end{eqnarray}
Here, the density of the single-particle Kohn-Sham orbitals are used to obtain the time dependent dipole moment $\overrightarrow{\mathcal{D}(t)}$, which is then Fourier transformed to attain the ACS $\sigma$.

In the examples which follow, the external electric field takes the form of a sine-squared pulse and is articulated in the following manner 
\begin{eqnarray}
\textbf{E}_{ext}(x,t) = A_{\ell}\sin^{2}{\Bigl[\omega_{env}t-\frac{\omega_{env}}{c}x\Bigr]}\sin{\Bigl[\omega t-\frac{\omega}{c}x\Bigr]}\hat{z} \nonumber \\ 
\label{Eext_sine}
\end{eqnarray}
Here, $A_{\ell}$, $\omega$, and $\omega_{env}$ are the intensity, radial frequency, and radial envelope frequency of the pulse. A linearly polarized external electromagnetic field is applied along the $\hat{z}$-direction and the electric field varies along the $x$-axis, where $\textbf{r}=(x,y,z)$. 

In the dipole approximation, we use the assumption of a spatially uniform external electric field to integrate the scalar potential $\phi_{ext}=\mathbf{E}\cdot\mathbf{r}$ into the Hamiltonian. Given we wish to describe a spatially varying external electric field, a different representation must be adopted. We instead use the gauge-free relation:

\begin{eqnarray}
	\textbf{E}_{ext}(\textbf{r},t) = -\frac{\partial\textbf{A}_{ext}}{\partial{t}} \label{gaugefree_E}
\end{eqnarray}
By then using a backward finite differencing technique, the discrete form of Eq.~\ref{gaugefree_E} for a time-step $\Delta{t}$ takes the form

\begin{eqnarray}
	\textbf{A}_{ext}(\textbf{r},t)=\textbf{A}_{ext}(\textbf{r},t-\Delta{t})-\Delta{t}\textbf{E}_{ext}(\textbf{r},t) \label{Aext_dscrt}
\end{eqnarray}

The external vector potential is then added to the Hamiltonian as in Eq. \ref{eom_dyn} along with a null scalar external potential.

\subsection*{Water}
We first study the ACS of a water molecule by applying a boost excitation along the $z$-direction.

The water molecule geometry contains an O-H bond length of $2.02$ a.u. and an H-O-H angle of $105.6\degree$, similar to that of del Puerto et al.~\cite{del2005real}. The domain size was a box of dimensions 24$\times$24$\times$ 24 a.u. and included absorbing boundary conditions (ABC) with a boundary layer of 4 a.u. at each side along each spatial dimension. A damping coefficient is carefully chosen to gradually transition the value of the orbital functions to zero at the boundary of the grid. The response of the water molecule required no more than $20$ femtoseconds (fs) and a time-step of $0.5$ attoseconds (as). 

\begin{figure}[!ht]
	\raisebox{-\height}{\includegraphics[width=\columnwidth]{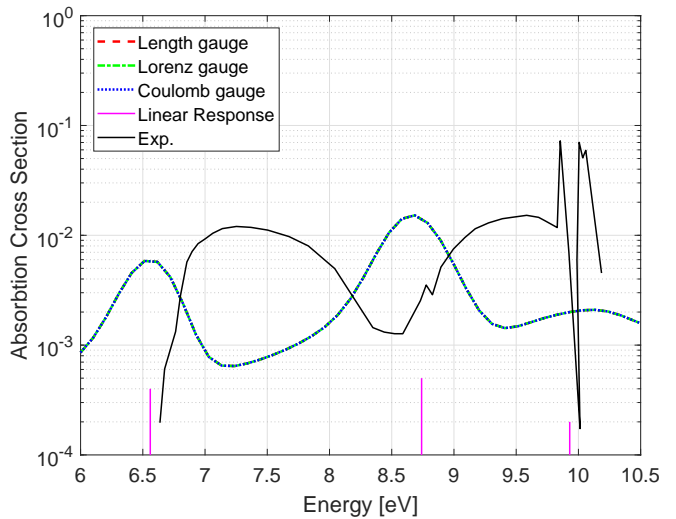}} 
	\caption{ACS of water for a boost excitation as a function of energy. Length (red), Lorenz (green), and Coulomb (blue) gauges attained using PARSEC. Comparison is made to the linear response (purple) and experimental (black) results of NWCHEM~\cite{Valiev20101477} and Yoshino \textit{et al}.~\cite{yoshino1996absorption} respectively.}.
	\label{fig:water_response} 
\end{figure}

As is apparent in figure~\ref{fig:water_response}, the ACS of the length, Lorenz, and Coulomb gauges align perfectly. The alignment of the Lorenz and Coulomb gauges suggests deep-lying invariance holds, at least on such small scales. 

\begin{figure}[!ht]
	\raisebox{-\height}{\includegraphics[width=\columnwidth]{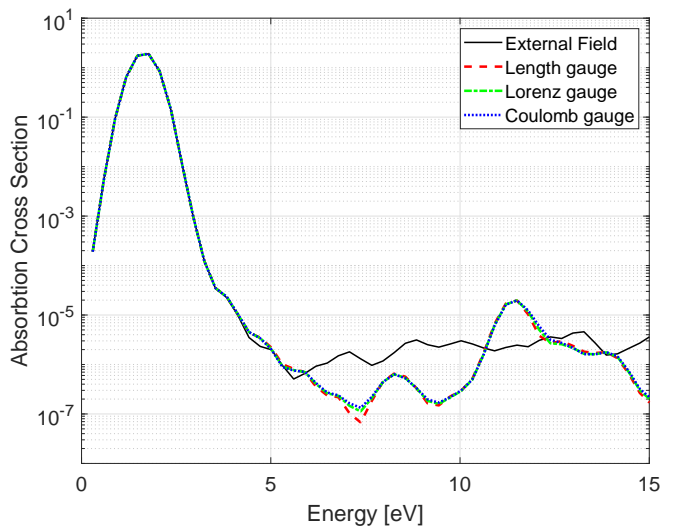}} 
	\caption{ACS of water for a sine squared pulse as a function of energy. Length (red), Lorenz (green), and Coulomb (blue) gauges attained using PARSEC.}
	\label{fig:water_sin2} 
\end{figure}
The results were broadened using a Gaussian width of $0.4$ eV, on the order of the vibrational energy of H$_{2}$O. The excitation energies obtained from the linear response procedure seem to be shifted from the experimental results of Yoshino \textit{et al}.~\cite{yoshino1996absorption}. Furthermore, the low energy peak of the time-propagation simulation, located at $6.56$ eV in figure~\ref{fig:water_response}, align perfectly with the linear response peak that was calculated with the NWCHEM package~\cite{Valiev20101477}. These results also seem to agree with the linear response results of del Puerto et al.~\cite{del2005real}. The intensity of the peaks located at $8.74$ eV and $9.93$ eV also nicely align with the aforementioned linear response calculation. 

Similar results are given in figure~\ref{fig:water_sin2} in the presence of the sine-squared pulse specified in Eq.~\ref{Eext_sine}. Clearly, H$_{2}$O is small enough for the various gauges to align as in figure~\ref{fig:water_response}.

\subsection*{Benzene}
We next demonstrate that the Benzene (C$_6$H$_6$) molecule ACS is also invariant to the choice of gauge. The benzene structure adopted here contains C$-$C and C$-$H bond lengths of $2.64$ a.u. and $2.05$ a.u., respectively. The domain was a box of dimensions 31$\times$31$\times$31 a.u. . ABCs are adopted, as before, but with an increased boundary layer of $5$ a.u. at each side along each dimension. The electronic structure was evolved in time for $60$ fs at a time-step of $1.0$ as using the same 4th-order Taylor expanded predictor-corrector scheme to insure properly converged dynamics.

\begin{figure}[!ht]
	\raisebox{-\height}{\includegraphics[width=\columnwidth]{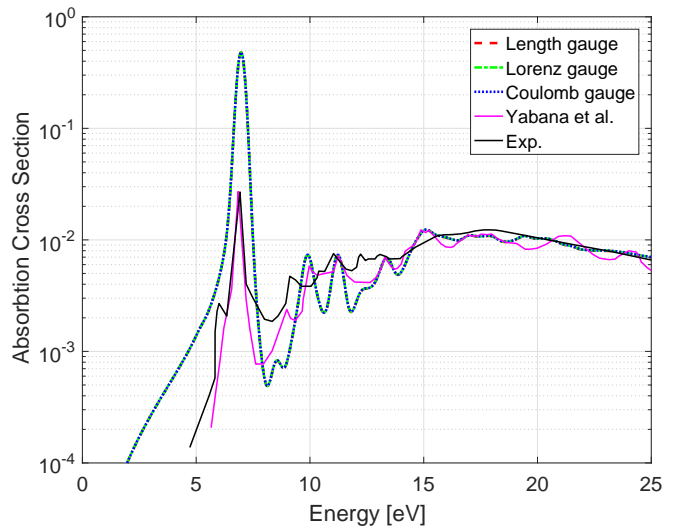}}
	\caption{ACS of benzene for an impulse excitation as a function of energy for the length (red), Lorenz (green), and Coulomb (blue) gauges. Real-time results of Yabana \textit{et al}.~\cite{yabana2006real} (purple) and the experimental results of Koch \textit{et al}.~\cite{koch1972optical} (black) are used for comparison.}
	\label{fig:benzene} 
\end{figure}

Figure~\ref{fig:benzene} depicts the ACS of benzene for the length, Lorenz, and Coulomb gauges, along with the experimental results of Koch \textit{et al}.~\cite{koch1972optical} and real-space, real-time results of Yabana \textit{et al}\cite{yabana2006real}. The different gauges all conform fairly well to one another and the experiment, in particular, they all agree on the  $\pi-\pi^{*}$ state transition at $6.9$ eV.

\begin{figure}[!ht]
	\raisebox{-\height}{\includegraphics[width=\columnwidth]{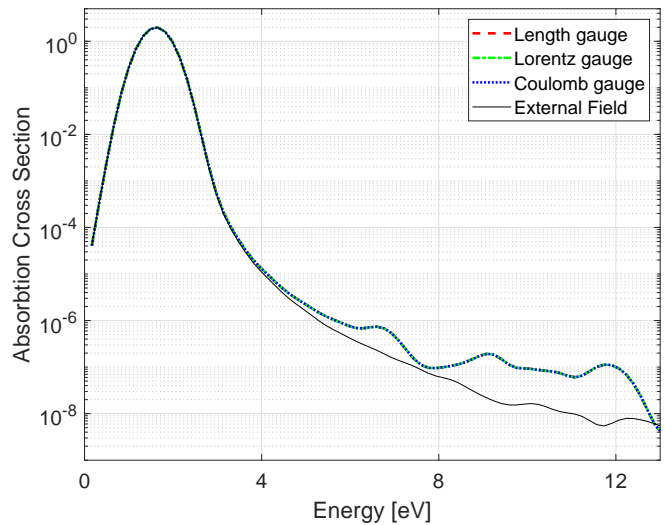}}
	\caption{ACS of benzene for a sine squared pulse as a function of energy for the length (red), Lorenz (green), and Coulomb (blue) gauges.}
	\label{fig:benzene_sine} 
\end{figure}

\subsection*{Cumulene}
We next calculated the response of a 1D carbon carbyne chain in the cumulene form,  H$_2$(=C=)$_{12}$H$_2$. This form of carbyne has a zero gap at the polymer limit\cite{calzolari2004ab} and can serve as a future candidate in studying retardation effects in significantly longer carbon chains.

\begin{figure}[!ht]
	\raisebox{-\height}{\includegraphics[width=\columnwidth]{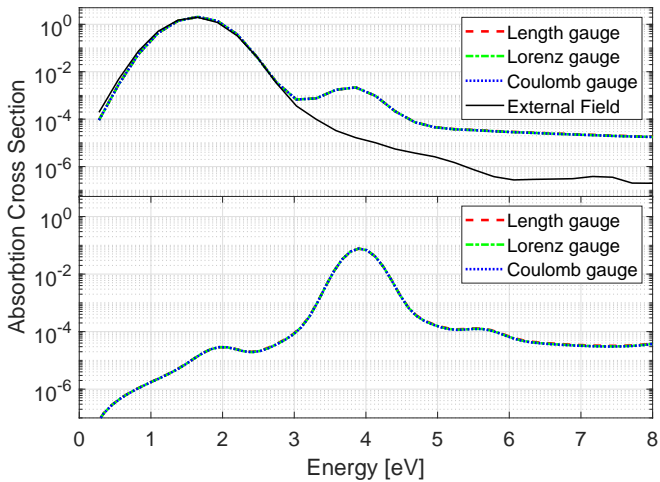}}
	\caption{ACS of 12C for a sine squared pulse (top) and impulse (bottom) as a function of energy for the length (red), Lorenz (green), and Coulomb (blue) gauges.}
	\label{fig:c12}
\end{figure}

The ACS of C$_{12}$H$_{4}$ is also shown to be gauge-invariant, as depicted in the bottom of figure~\ref{fig:c12}. The elongated component of the carbon chain contains a sharp excitation peak at $\sim4$ eV. In generating these results, a grid of $22$ a.u. along the cross sections and $48$ a.u. in the elongated direction was adopted. Once again, ABCs with a boundary layer of $5$ a.u. along each side of the box were used to remove any reflections resulting from nonzero orbital functions at the boundary. The ACS was attained by propagating the single-particle orbitals for over $90$ fs at a time-step of $0.5$ as. .
Unlike the previous molecular structures, it took much longer for the lower frequency peaks to appear after the boost excitation was applied. 

In a similar manner, the top of figure~\ref{fig:c12} depicts the excitation spectra of the length, Lorenz, and Coulomb gauges in the presence of a $800$ nm external electromagnetic field. The additional peak located at $\sim4.0$ eV is characteristic of a quantum resonance. The location of the resonance comes as no surprise in observing the analogous $4$ eV excitation peak of the aforementioned impulse response in the bottom of figure~\ref{fig:c12}. As in the impulse response, the gauge-invariance is once again numerically demonstrated.

\section{Discussion \& Summary}

In this work we have described how the Lorenz and Coulomb gauges can be formally introduced, as an approximation, into the the TDKS Hamiltonian. We have also shown a numerically efficient way to evaluate the integral expressions of the retarded potentials in both gauges. This approach was demonstrated in real-space for small molecules, such as water, benzene, and cumulene (C$_{12}$H$_{4}$), by propagating single-particle orbitals in real-time. We have further demonstrated that the various gauges aligned with the results of the length gauge for such molecular structures. While this result may be trivial for small systems with weak retardation effects, such a formal approach can be easily extended to multi-domain scenarios of interacting small systems.

The integration of faster propagators and better parallelization schemes will enable the calculation of systems large enough to have more significant retardation effects, while still residing in the weak magnetic fields limit. The simultaneous consideration of two gauges can help verify whether gauge-invariance is still achievable in practice. Reaching such larger scales can further enable the consideration of nanodevices, such as nanorectennas~\cite{donchev2014nanorect,slepyan2017nanorect}, which were, to our knowledge, never studied numerically from first principles. 

It should be emphasized that the velocity gauges studied thus far bring about a much richer physical characterization of the electronic structure then their length gauge counterpart. The induced magnetic fields essentially play the role of allowing classical solitons, created by the dynamical behavior of a given local regime of the electronic structure, to affect another distant local regime in a causal manner. This classical effect can play an important role for large system sizes, even at a fraction of the external electromagnetic field's wavelength. 

Induced magnetic fields can play a prior role in other components of the Kohn-Sham Hamiltonian, particularly in considering the non-collinear form of the spin. The induced magnetic field within the Stern-Gerlach term of the Kohn-Sham-Pauli equation\cite{sharma2007first} may be dramatically altered in characterizing the spin dynamics of electronic structures, particularly if they exceed the exchange-correlation magnetic fields defined by the difference in spin-up and spin-down exchange-correlation scalar potentials. 

In this work, the Kohn-Sham Hamiltonian was extended to include retarded potentials, as opposed to the sole consideration of the electrostatic Hartree potential within TDDFT. In incorporating the vector potential into the Kohn-Sham equation, TDDFT is effectively extended to incorporate induced electrodynamic fields. The nontrivial integral expressions were evaluated using an accelerative scheme, making the study of large electronic structures tractable. The Lorenz and Coulomb gauge-fixing conditions were studied and gauge-invariance was exemplified in the weak magnetic field limit. It was further proposed that, in studying dynamical properties, adopting the Lorenz gauge may be formally simpler than its Coulomb gauge counterpart. 

An interesting prospect arises in considering alternative gauges, particularly, the more natural characterization of the dynamical behavior of scalar and vector potentials within the Lorenz gauge. In considering the history of densities within the exchange-correlation functionals, the intrinsically causal structures of the Lorenz gauge-fixed potentials could make the articulation of dynamical many-body effects formally simpler, removing the need to deal directly with the unintuitive nature of the transverse vector potential. 

\begin{acknowledgments}
We thank Dr. Gregory Slepyan of Tel Aviv University (TAU) for helpful discussions on the subject of applying alternative gauge-fixing conditions to TDDFT. 
This research was supported by grants from the United States-Israel Binational Science Foundation (BSF), Jerusalem, Israel, under BSF grant numbers 2014426 \& 2018182.
This research was also partially funded by EU H2020, under the Project TERASSE, H2020-MSCA-RISE, 823878
\end{acknowledgments}

\appendix
\section{Coulomb Gauge - Integral Expressions}
Integral expressions resulting from the Coulomb gauge fixing condition are derived. This appendix is provided for the purpose of completeness. The gauge-free relations are
\begin{eqnarray}
\textbf{E}=-\nabla{v}-\frac{1}{c}\frac{\partial\textbf{A}}{\partial{t}} \label{gauge_rltn1}
\end{eqnarray}
\begin{eqnarray}
\textbf{B}=\nabla\times\textbf{A} \label{gauge_rltn2}
\end{eqnarray}
Adopting the first of these gauge free relations to Gauss's Law, and applying the Coulomb gauge-fixing condition $\nabla\cdot\textbf{A}=0$, one can attain the well-known Poisson equation for the scalar potential
\begin{eqnarray}
\nabla\cdot\textbf{E}=-\nabla\cdot\Bigl(\nabla{v}+&&\frac{1}{c}\frac{\partial\textbf{A}}{\partial{t}}\Bigr)=n \nonumber \\
-\nabla^{2}{v}+\frac{1}{c}\frac{\partial}{\partial{t}}(\nabla\cdot&&\textbf{A})=n \nonumber \\
\nabla^{2}{v}=-n&& \label{gaussc}
\end{eqnarray}
Similarly, one can arrive at the D'Alembert equation for the vector potential using Ampere's Law and Eq.~\ref{gauge_rltn2}
\begin{eqnarray}
\nabla\times\textbf{B}&&=\nabla\times\nabla\times\textbf{A}_{T}=\nabla(\nabla\cdot\textbf{A}_{T})-\nabla^{2}\textbf{A}_{T} \nonumber \\ 
-\nabla^{2}\textbf{A}_{T}&&=\textbf{j}-\Bigl(\frac{1}{c}\frac{\partial^{2}\textbf{A}_{T}}{\partial{t}^{2}}+\frac{1}{c}\frac{\partial^{2}\textbf{A}_{L}}{\partial{t}^{2}}+\frac{\partial}{\partial{t}}(\nabla{v})\Bigr) \label{amperec1}
\end{eqnarray}
Here, the transverse and longitudinal vector potentials have been separated $\textbf{A}=\textbf{A}_{T}+\textbf{A}_{L}$. The curl of any vector is transverse by definition, suggesting $\nabla\times\textbf{A}=\nabla\times\textbf{A}_{T}$. Additionally, by the Coulomb gauge, $\nabla\cdot\textbf{A}=0$. In rearranging Eq.~\ref{amperec1}, we arrive at
\begin{eqnarray}
\frac{1}{c}\frac{\partial^{2}\textbf{A}_{T}}{\partial{t}^{2}}-\nabla^{2}\textbf{A}_{T}&&=\textbf{j}-\frac{\partial}{\partial{t}}(\nabla{v})=\textbf{j}-\textbf{j}_{L} \nonumber \\
\Box\textbf{A}_{T}&&=\textbf{j}_{T} \label{amperec2}
\end{eqnarray}
Eq.~\ref{gaussc} and Eq.~\ref{amperec2} serve as the differential equations defined for the Coulomb gauge, where the transverse current density $\textbf{j}_{T}$ is attained via Helmholtz decomposition, as formulated in eq.~\ref{helmholtz_v0}. In assuming a Green's function exists for each of these individual operators $\mathcal{P}$, 
\begin{eqnarray}
\mathcal{P}G(\textbf{r},t,\textbf{r}',t')=\delta(\textbf{r}-\textbf{r}')\delta(t-t') \label{greensD}
\end{eqnarray}
One arrives at the well known expressions specified in Eq.~\ref{coulomb_gaugeS}-\ref{coulomb_gauge}.

\section{Lorenz Gauge - Integral Expressions}
Integral expressions resulting from the Lorenz gauge-fixing condition are derived. This appendix is provided for the purpose of completeness and can be found in any standard textbook. Using the gauge free relations of Eq.~\ref{gauge_rltn1}-\ref{gauge_rltn2}, Gauss's Law can be used to define the scalar potential
\begin{eqnarray}
\nabla\cdot\textbf{E}=-\nabla\cdot\Bigl(\nabla{v}+&&\frac{1}{c}\frac{\partial\textbf{A}}{\partial{t}}\Bigr)=n \nonumber \\
-\nabla^{2}{v}+\frac{1}{c}\frac{\partial}{\partial{t}}\Bigl(\frac{\partial{v}}{\partial{t}}&&\Bigr)=n \nonumber \\
\Box{v}=n&& \label{gaussl}
\end{eqnarray}
Here, the Lorenz gauge fixing condition $\nabla\cdot\textbf{A}=\partial{v}/\partial{t}$ is applied to the first line. One can similarly derive the expression for the vector potential starting from Eq.~\ref{amperec1}
\begin{eqnarray}
-\nabla^{2}\textbf{A}_{T}=\textbf{j}-&&\Bigl(\frac{1}{c}\frac{\partial^{2}\textbf{A}}{\partial{t}^{2}}-\nabla(\nabla\cdot\textbf{A}_{L})\Bigr) \nonumber \\
\frac{1}{c}\frac{\partial^{2}\textbf{A}}{\partial{t}^{2}}-\nabla^{2}\textbf{A}_{T}&&=\textbf{j}+\nabla^{2}\textbf{A}_{L}+\nabla\times\nabla\times\textbf{A}_{L}  \nonumber \\ 
&&\Box\textbf{A}=\textbf{j} \label{amperel1}
\end{eqnarray}
Here, the divergence of the vector potential is purely longitudinal $\nabla\cdot\textbf{A}=\nabla\cdot\textbf{A}_{L}$. It then naturally follows that $\nabla\times\textbf{A}_{L}=0$. The integral expressions (Eq.~\ref{lorentz_gaugeS}-\ref{lorentz_gauge}) corresponding to Eq.~\ref{gaussl}-\ref{amperel1} can then be attained via Eq.~\ref{greensD}.

\end{document}